\documentclass[aps,twocolumn,floats,pre,nofootinbib]{revtex4}
\usepackage{graphics,graphicx,epsfig}
\usepackage{amssymb,color}
\usepackage{epsf,epstopdf,wrapfig}
\usepackage {amsmath}

\newcommand{\beq}{\begin{equation}}
\newcommand{\eeq}{\end{equation}}
\newcommand{\beqn}{\begin{eqnarray}}
\newcommand{\eeqn}{\end{eqnarray}}

\begin{document}

\title{Searching for sequence features that control DNA flexibility}

\author{Yaojun Zhang,$^1$  Aakash Basu,$^2$ Taekjip Ha,$^{2,3,4,5}$ and William Bialek$^{1,6}$}

\affiliation{$^1$Joseph Henry Laboratories of Physics and Lewis--Sigler Institute for Integrative Genomics, Princeton University, Princeton, NJ 08544\\
$^2$Department of Biophysics and Biophysical Chemistry, Johns Hopkins University School of Medicine, Baltimore, MD 21205\\ 
$^3$Department of Biophysics, Johns Hopkins University, Baltimore, MD 21218\\ 
$^4$Department of Biomedical Engineering, Johns Hopkins University, Baltimore, MD 21205\\ 
$^5$Howard Hughes Medical Institute, Baltimore, MD 21205\\ 
$^6$Initiative for the Theoretical Sciences, The Graduate Center, City University of New York, 365 Fifth Ave, New York, NY 10016}

\begin{abstract}
Modern genomics experiments measure functional behaviors for many thousands of DNA sequences. We suggest that, especially when these sequences are chosen at random, it is natural to compute correlation functions between sequences and measured behaviors. In simple models for the dependence of DNA flexibility on sequence, for example, correlation functions can be interpreted directly as interaction parameters. Analysis of recent experiments shows that this is surprisingly effective, leading directly to extraction of distinct features for DNA flexibility and predictions that are as accurate as more complex models. This approach follows the conventional use of correlation functions in statistical physics and connects the search for relevant DNA sequence features to the search for relevant stimulus features in the analysis of sensory neurons.
\end{abstract}

\date{\today}

\maketitle

In physics we often use correlation functions to characterize the behavior of a system,  and many experimentally measurable quantities are related directly to these correlation functions.  As examples, the diffusion constant of a particle is an integral over the correlation function of its velocity, the X--ray diffraction pattern of a material is the Fourier transform of the correlation function of density fluctuations \cite{chaikin+lubensky_95}, and scattering amplitudes for elementary particles are correlation functions in the underlying quantum field theory that describes their interactions \cite{peskin+schroeder_95}.    It has taken longer for this language of correlation functions to permeate the analysis of living systems.  

In analyzing how single neurons respond to their inputs, it is conventional to compute the correlation between the continuous inputs and the discrete sequence of action potentials or spikes at the output \cite{deboer+kuyper_68,rieke+al_97,dayan+abbott_01}; this ``triggered correlation'' seems to have been inspired more by ideas of systems identification in engineering than correlation functions in physics \cite{wiener_58}.  It eventually was  realized that this approach could be generalized to higher order correlations, allowing the identification of multiple relevant input features in triggering a spike \cite{ruyter+bialek_88,bialek_ruyter_05}.  In these applications, it is important that the inputs can be chosen from appropriate ensembles.  More recently, correlation functions have emerged as central to the analysis of collective behavior in animal groups, much in the original spirit of their use to analyze experiments in condensed matter \cite{cavagna+al_18}.  Here we consider the use of correlation functions to analyze experiments on the mechanics of randomly chosen DNA sequences \cite{basu+al_20a}.

The key step in using correlation functions to analyze neural responses was to shift from measuring responses to particular, carefully chosen sensory stimuli \cite{hubel+wiesel_62} to an unbiased exploration of many more stimuli chosen randomly from some well understood distribution.  As an example, if a neuron integrates for $\sim 100\,{\rm msec}$, then recording neural activity in response to one hour of continuous random inputs is equivalent to sampling $\sim 3\times 10^4$ different stimuli.  Long before the genomic revolution brought the term into common use, this approach thus achieved ``high throughput.'' 

To make the discussion concrete, we consider DNA sequences $\{S_{\rm i}^\alpha\}$, where $S_{\rm i}^\alpha = 1$ if the base at site $\rm i$ is of type $\alpha$, and $S_{\rm i}^\alpha = 0$ otherwise.  The index ${\rm i} = 1,\, 2,\, \cdots ,\, N$, where $N$ is the length of the sequences we are studying, and $\alpha = 1,\,2,\,3,\,4$, corresponding to A, T, C, G.  If we choose sequences at random from the uniform distribution, we have
$\langle S_{\rm i}^\alpha \rangle = 1/4$ and
\begin{equation}
\langle S_{\rm i}^\alpha S_{\rm j}^\beta \rangle = \delta_{\rm ij}\delta^{\alpha\beta} (1/4) + (1-\delta_{\rm ij})(1/4)^2 ,
\end{equation}
which means that the connected correlations are
\begin{eqnarray}
\langle S_{\rm i}^\alpha S_{\rm j}^\beta \rangle_c &\equiv& \langle S_{\rm i}^\alpha S_{\rm j}^\beta \rangle - 
\langle S_{\rm i}^\alpha \rangle\langle S_{\rm j}^\beta \rangle \nonumber\\
&=& \langle (S_{\rm i}^\alpha -1/4)(S_{\rm j}^\beta -1/4)\rangle \nonumber\\
&=& \delta_{\rm ij} (1/4)\left( \delta^{\alpha\beta} - 1/4\right) .
\end{eqnarray}
We can go on to compute higher order correlations, which will be relevant below; details  are  in Appendix \ref{details}.

\begin{figure}
\centerline{\includegraphics[width = 0.9\linewidth]{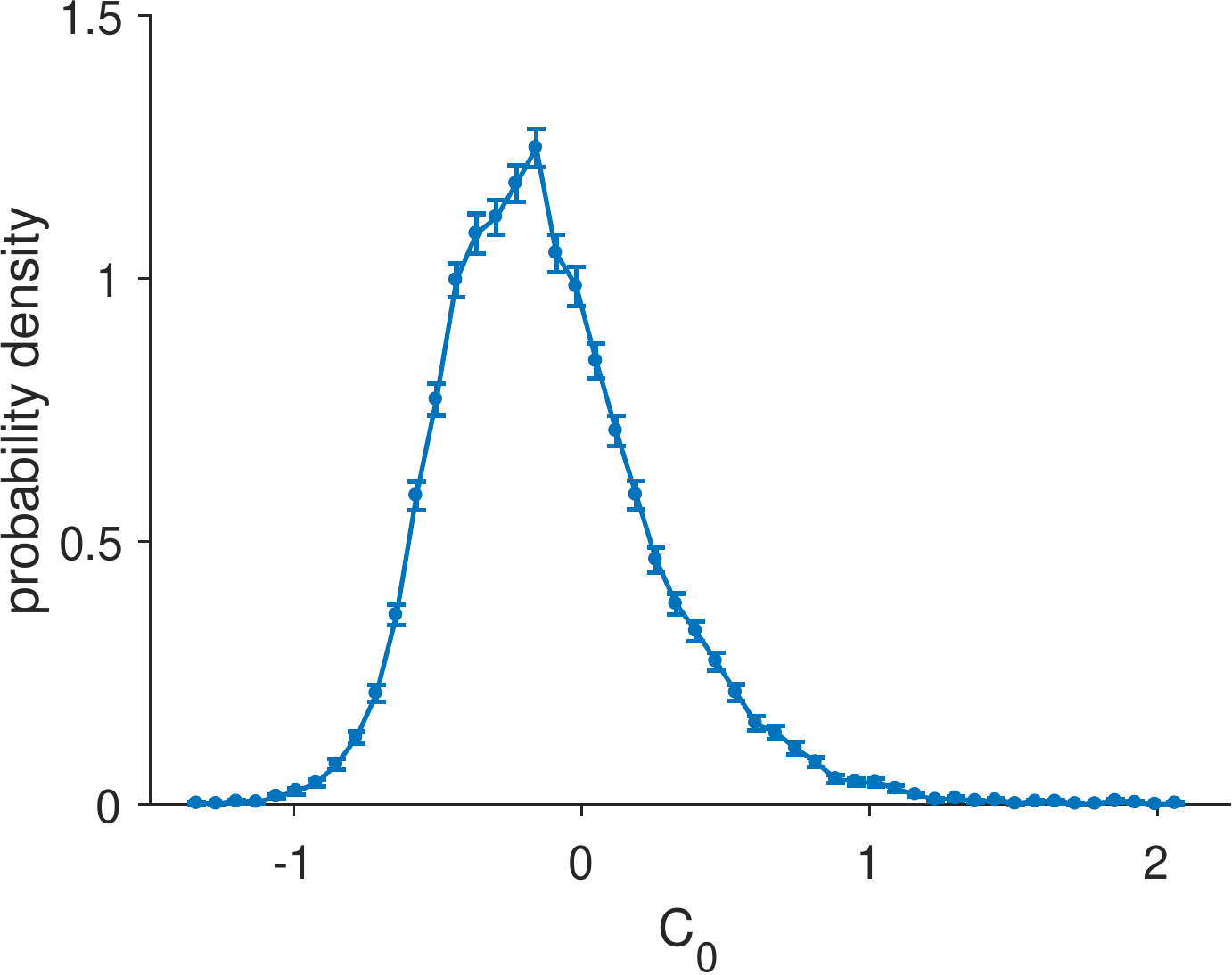}}
\caption{The distribution of the intrinsic cyclizability $C_0$ across the $\sim 10^4$ random sequences in the experiment of  Ref \cite{basu+al_20a}.  Mean and standard deviation across random halves of the data. \label{PC0}}
\end{figure}

Recent experiments have chosen $M = 12,472$ random sequences from the uniform distribution and estimated the intrinsic flexibility of these sequences by measuring the probability that they close on themselves into a loop \cite{basu+al_20a}.  In detail, randomly chosen sequences of length $N=50$ were flanked by fixed double stranded adapters and complementary overhangs, and immobilized on a bead.  The looping reaction was initiated by changing solution conditions, and after a fixed time the unlooped molecules were degraded by an enzyme that only attacks free ends.  The remaining population of looped molecules was sequenced and compared with the original ensemble; cyclizability was defined as the log ratio of probabilities for finding sequences in the looped vs control ensembles. Observations on a small number of sequences show that this measure  correlates very well with direct measurements of flexibility on single molecules.   The measured cyclizability depends periodically on the location of the bead attachment, and the intrinsic cyclizability $C_0$ was defined as the mean over this variation. The distribution of $C_0$ across the sequences is shown in Fig \ref{PC0}.

\begin{figure}[b]
\centerline{\includegraphics[width = 0.95\linewidth]{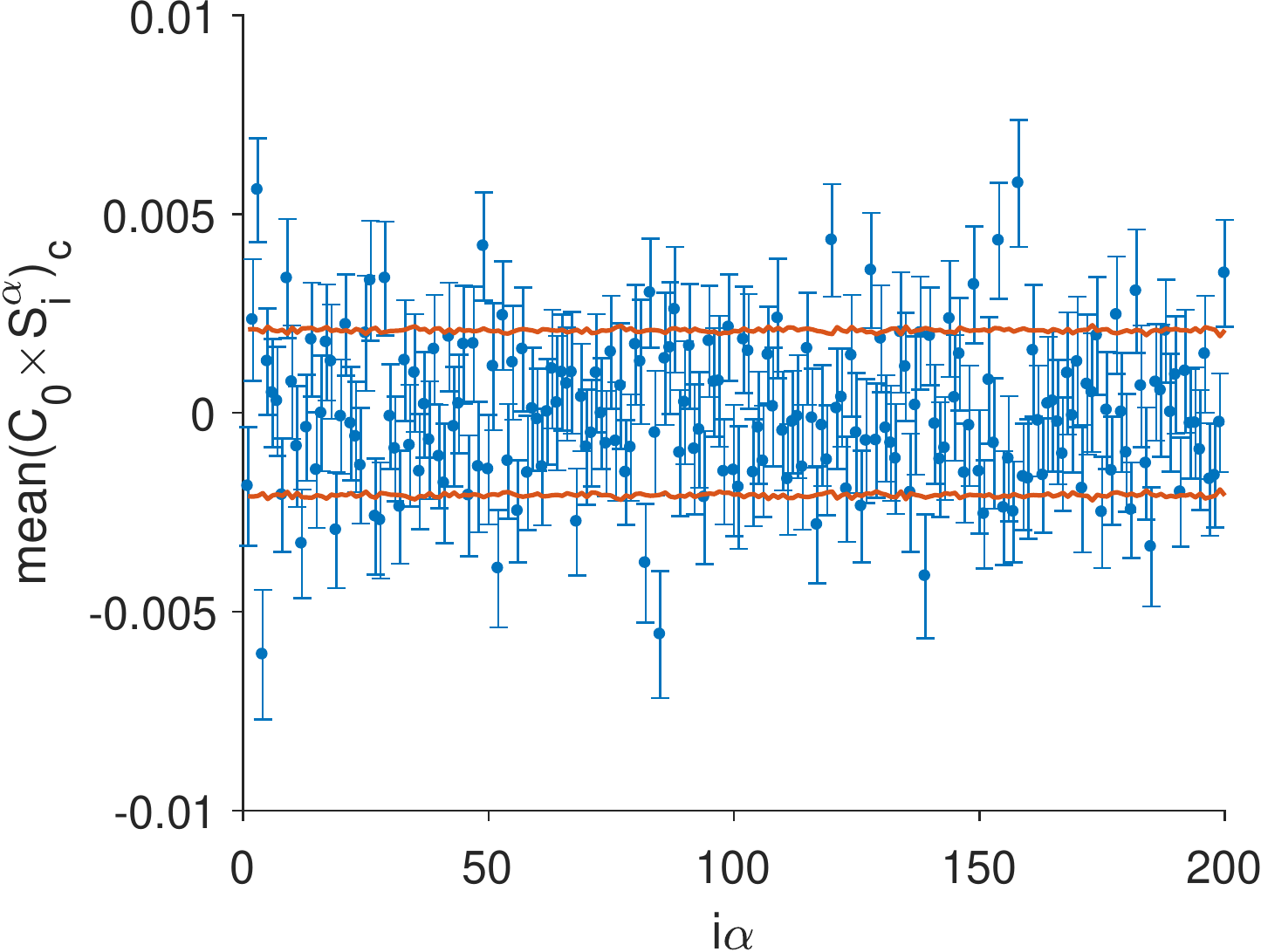}}
\caption{The connected correlation function $\langle C_0 S_{\rm i}^\alpha\rangle_c$. Blue points: Mean and standard deviation across random halves of the data.  Red lines: $\pm$ one standard deviation across random halves of shuffled data.  \label{corr1}}
\end{figure}

The simplest model for how the cyclizability depends on sequence is linear,
\begin{equation}
C_0 = \langle C_0\rangle + \sum_{{\rm j},\beta} W_{\rm j}^\beta \left( S_{\rm j}^\beta - 1/4\right) ,
\label{PWM}
\end{equation}
where $W_{\rm i}^\alpha$ is analogous to the position weight matrices that appear in models of transcription factor binding \cite{berg+vonhippel_87,stormo_00,kinney+al_07}.   Without loss of generality we can set $\sum_\beta W_{\rm j}^\beta =0$ at every site $\rm j$.  If this model is correct, then we can isolate the elements of $W$ by computing a (connected) correlation function, averaging over random sequences, 
\begin{eqnarray}
\langle C_0 S_{\rm i}^\alpha\rangle_c &\equiv&
\langle \left(C_0 -\langle C_0\rangle \right)\left( S_{\rm i}^\alpha - \langle S_{\rm i}^\alpha \rangle \right)\rangle \nonumber\\
 &=& \sum_{{\rm j},\beta} W_{\rm j}^\beta \langle S_{\rm j}^\beta   S_{\rm i}^\alpha \rangle_c \nonumber\\
&=& {1\over 4} \sum_{{\rm j},\beta} W_{\rm j}^\beta\delta_{\rm ij} (\delta^{\alpha\beta} - 1/4)=  {1\over 4}  W_{\rm i}^\alpha  .
\end{eqnarray}
We show this correlation function, computed from the data, in Fig \ref{corr1}.  The results are consistent with $\langle C_0 S_{\rm i}^\alpha\rangle_c =0$, suggesting that there is no linear term in the dependence of $C_0$ on the sequence.

If Equation (\ref{PWM}) doesn't work, because the data are consistent with $W = 0$, the next simplest model is
\begin{equation}
C_0 = \langle C_0\rangle + {1\over 2}\sum_{{\rm kl},\gamma\delta} J_{\rm kl}^{\gamma\delta}( S_{\rm k}^\gamma - 1/4)(S_{\rm l}^\delta - 1/4) .
\label{2ndModel}
\end{equation}
As shown in Appendix \ref{details},  any site diagonal term $J_{\rm ii}^{\alpha\beta}$ in the matrix $J$ can be rewritten as a weight $W_{\rm i}^\alpha$ in the linear model, so we can set these terms to zero.  We also can set $\sum_\beta J_{\rm ij}^{\alpha\beta} =0$, since $\sum_\alpha S_{\rm i}^\alpha = 1$.
Now we want to compute the correlation function 
\begin{equation}
\langle C_0 S_{\rm i}^\alpha S_{\rm j}^\beta\rangle_c 
\equiv
 \langle \left( C_0 -\langle C_0\rangle \right)  ( S_{\rm i}^\alpha -\langle S_{\rm i}^\alpha \rangle ) ( S_{\rm j}^\beta -\langle S_{\rm j}^\beta \rangle ) \rangle ,  \label{C2}
\end{equation}
and we find (see Appendix \ref{details} for details) that 
\begin{equation}
\langle C_0 S_{\rm i}^\alpha S_{\rm j}^\beta\rangle_c =  {1\over {16}} J_{\rm ij}^{\alpha\beta}  .
\label{trig_corr}
\end{equation}
As in the case of the linear model,  computing correlation functions over random sequences directly recovers the underlying interaction parameters.  

\begin{figure}
  \centering
  \includegraphics[width=.9\linewidth]{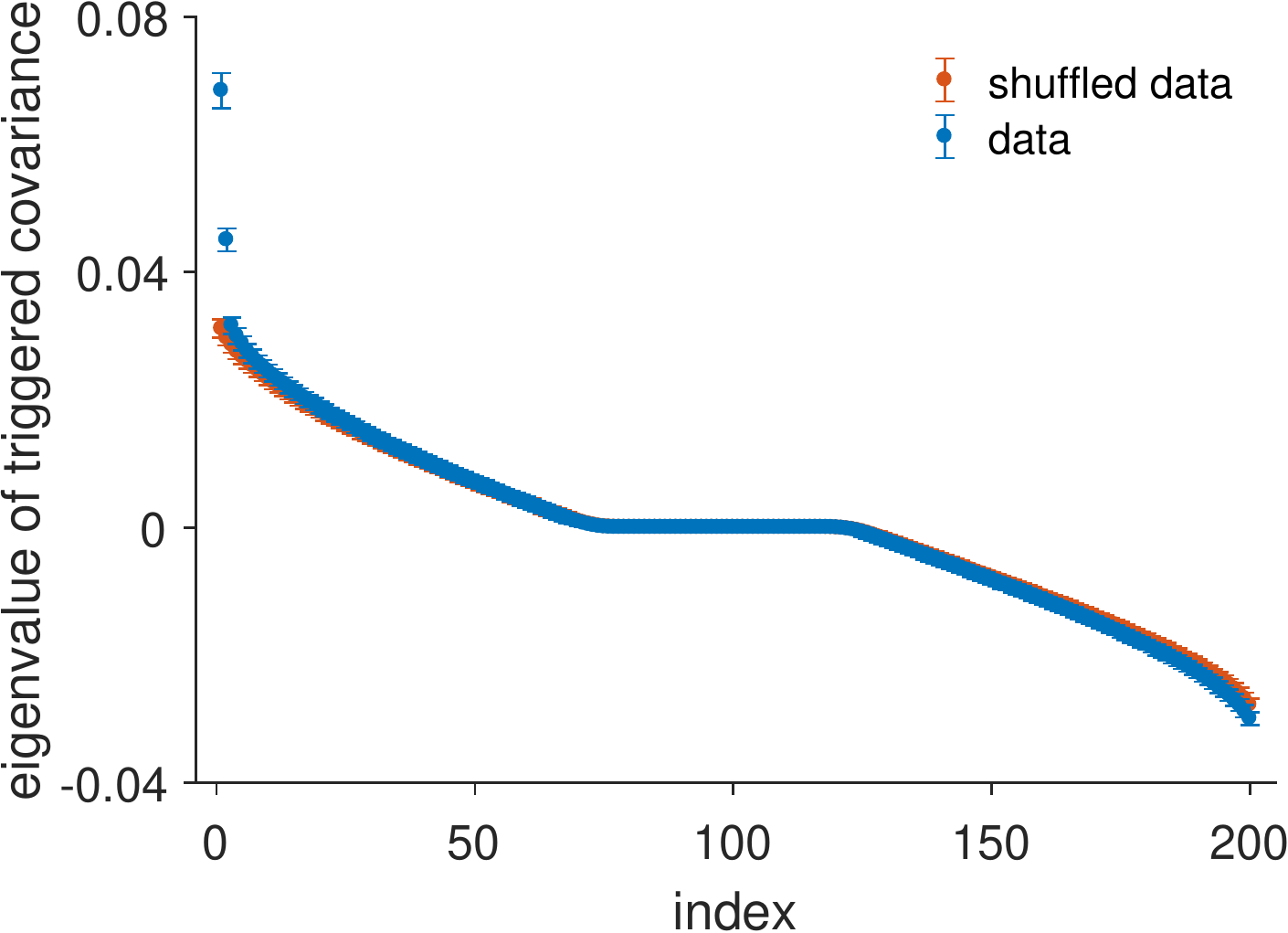}
\par\bigskip
  \includegraphics[width=.95\linewidth]{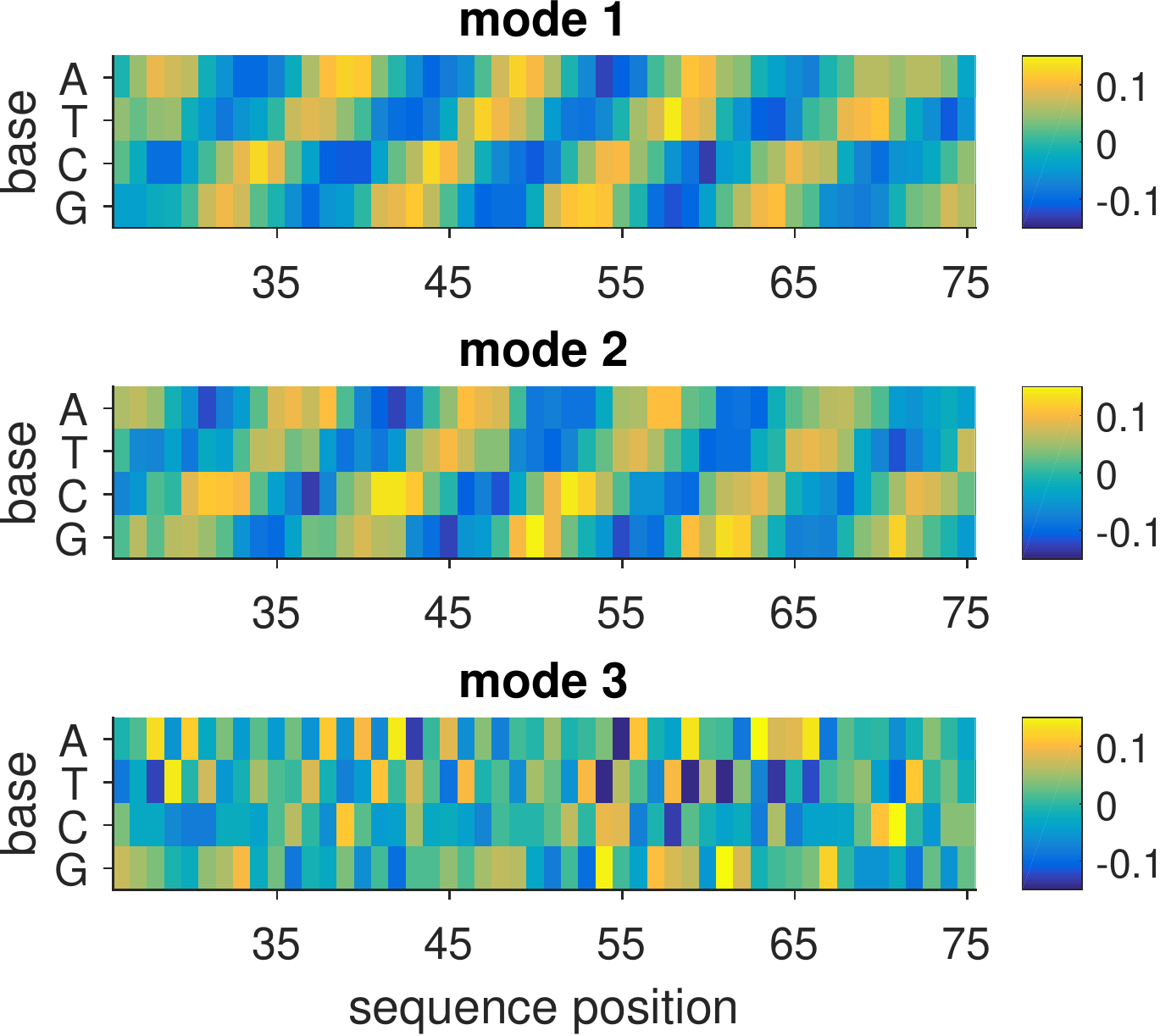}
\caption{Eigvenvalues and eigenvectors of the correlation matrix $\langle C_0 S_{\rm i}^\alpha S_{\rm j}^\beta\rangle_c$, Eq~(\ref{eigs}). (top) Eigenvalues  $\lambda_{\rm n}$ (blue) compared with results from shuffled data (red); points are means and error bars are standard deviations across randomly chosen halves of the sequences. (bottom) Leading eigenvectors $w_{\rm i}^\alpha ({\rm n})$, for ${\rm n}=1,\,2,\,3$. Scale is set by normalization, Eq~(\ref{norm}).\label{eigenmodes}}
\end{figure}

We emphasize that this correlation function is a matrix:  we can combine the indices $({\rm i},\alpha) \rightarrow \mu$ and $({\rm j},\beta) \rightarrow \nu$ so that  $\langle C_0 S_{\rm i}^\alpha S_{\rm j}^\beta\rangle_c \rightarrow M_{\mu\nu}$.  This construction thus is analogous to the spike--triggered covariance matrix in the analysis of neural responses \cite{bialek_ruyter_05}.  We search for further simplification by analyzing  eigenvalues and eigenvectors,
\begin{equation}
\langle C_0 S_{\rm i}^\alpha S_{\rm j}^\beta\rangle_c = \sum_{\rm n} \lambda_{\rm n} w_{\rm i}^\alpha ({\rm n}) w_{\rm j}^\beta ({\rm n}) ;
\label{eigs}
\end{equation}
it will be important that eigenvectors are orthonormal,
\begin{equation}
\sum_{{\rm i,\alpha} } w_{\rm i}^\alpha ({\rm n}) w_{\rm i}^\alpha ({\rm m})  = \delta_{\rm nm}.
\label{norm}
\end{equation}

We estimate the correlation function $\langle C_0 S_{\rm i}^\alpha S_{\rm j}^\beta\rangle_c $ from the data, and then diagonalize.  In Fig~\ref{eigenmodes} (top) we show the spectrum of eigenvalues $\{\lambda_{\rm n}\}$, in rank order, and compare with data that have been shuffled to break any correlations between sequence and flexibility.  We first notice that in both the real and shuffled data there are some true zero eigenvalues.  These arise because  we have $\sum_\alpha S_{\rm i}^\alpha = 1$ at each site $\rm i$, by definition.   In the shuffled data we see a spreading of the eigenvalues, which arises because we are estimating the correlation function from a finite sample   \cite{potters+bouchaud_20,note1}.  But in the real data there are at least two ``modes'' that stand out from this background.  

The eigenvectors $ w_{\rm i}^\alpha ({\rm n})$ have the same structure as position weight matrices, and pick out modes of sequence variation.  Figure~\ref{eigenmodes} (bottom) shows the three leading modes.  We note that the first two have a clear structure, while the third---with its eigenvalue less clearly distinguished from the background noise in Fig~\ref{eigenmodes}---seems almost random. The first two modes show an approximate ten base periodicity, consistent with the pitch of the double helix, and are close to being a quadrature pair.

We expect  eigenvectors to form exact quadrature pairs  if  $ J$ is invariant to translations along the chain,
\begin{equation}
J_{\rm ij}^{\alpha\beta} = J_{{\rm j}-{\rm i}}^{\alpha\beta}.
\end{equation}
If we think of $J_{\rm ij}^{\alpha\beta}$ as an interaction between the bases as positions $\rm i$ and $\rm j$, then translation invariance is the statement that interactions depend on separation but not on absolute position. We can impose translation invariance by estimating $ J$ from the data using the correlation function in Eq (\ref{trig_corr}) and then replacing each matrix element by the average of all elements with the same value of ${\rm j} - {\rm i}$,
\begin{equation}
J_{{\rm i},{\rm j}>{\rm i}}^{\alpha\beta}
\rightarrow {1\over{N-{\rm j}+ {\rm i}}}\sum_{{\rm k}=1}^{N-{\rm j}+ {\rm i}} J_{{\rm k},{\rm k}+{\rm j}-{\rm i}}^{\alpha\beta}. \label{invariance}
\end{equation}
As detailed in Appendix \ref{clean}, the eigenvalues of this ``cleaned'' matrix stand out from the shuffled background with higher signal to noise ratio, both at large positive and large negative values;  the eigenvectors are more clearly periodic; and eigenvalues come in degenerate pairs.  We note that by imposing translation invariance, the number of independent parameters in the $ J$ matrix is reduced from $\sim15000$ to $\sim600$, which significantly raises the signal to noise ratio of the inferred $J$ matrix.

\begin{figure}[t]
\centerline{\includegraphics[width = \linewidth]{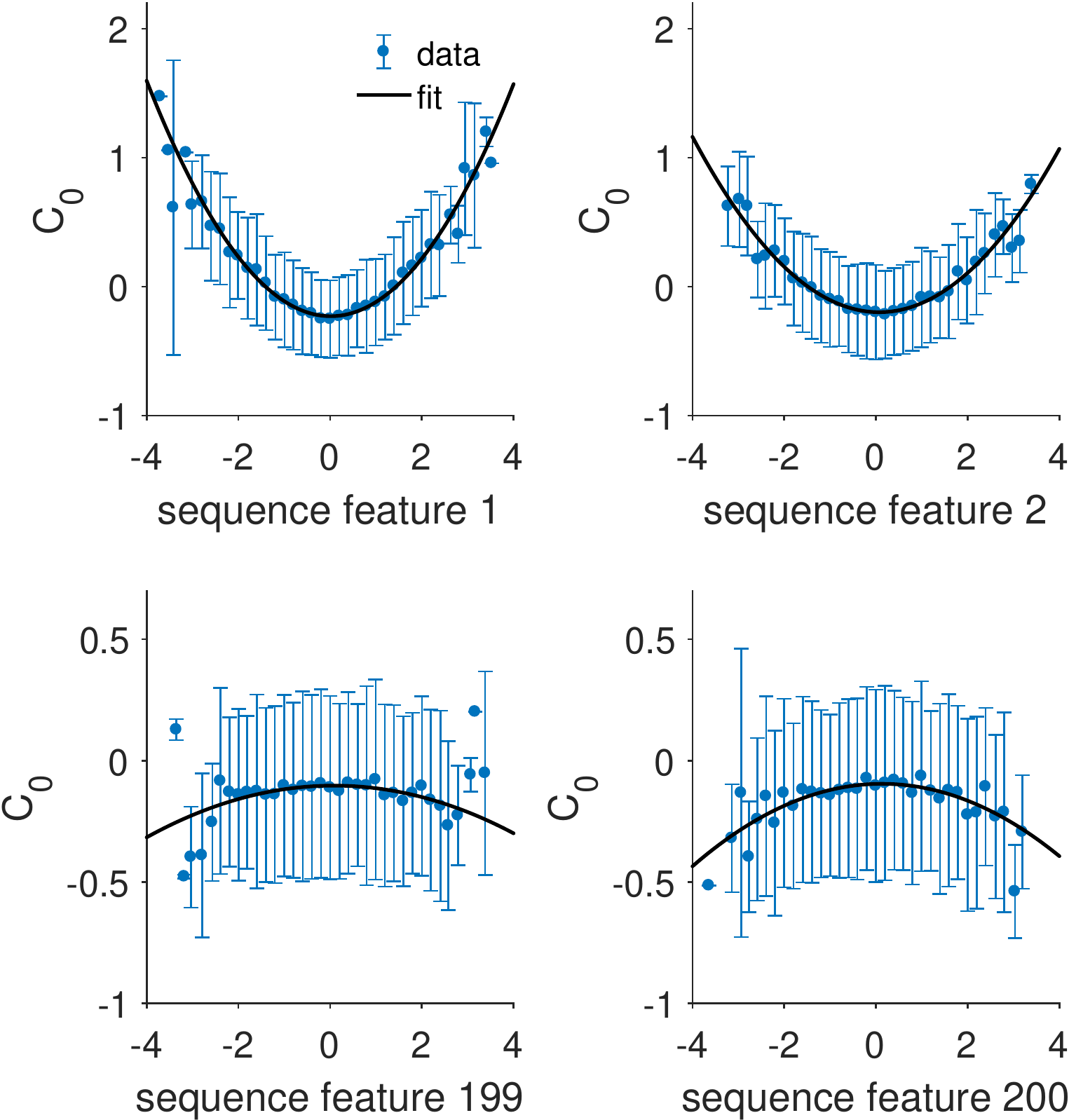}}
\caption{Cyclizability as a function of sequence features, from Eq (\ref{feat}).  Error bars are standard deviations across random splittings of the data into 50/50 training/test sets. Lines are quadratic fits, as  expected from Eq (\ref{2ndModel}).  Each feature is measured in units of its standard deviation across the ensemble of sequences. \label{Cvsf}}
\end{figure}

We can decompose the sequence variations into modes defined by the eigenvectors, forming sequence features
\begin{equation}
f_{\rm n} = \sum_{{\rm i},\alpha} w_{\rm i}^\alpha ({\rm n}) S_{\rm i}^\alpha .
\label{feat}
\end{equation}
In Fig \ref{Cvsf} we show the dependence of the cyclizability $C_0$ on the $f_{\rm n}$ at the extremes of the spectrum. To avoid overfitting we estimate $\langle C_0 S_{\rm i}^\alpha S_{\rm j}^\beta\rangle_c $  and hence the eigenvectors $w_{\rm i}^\alpha ({\rm n})$ from half of the sequences, and then probe $C_0$ vs $f_{\rm n}$ in the other half of the data. The mean behavior is almost perfectly quadratic along each feature, as predicted from Eq (\ref{2ndModel}), and consistent with the absence of any linear correlation between sequence and $C_0$. 

\begin{figure}[b]
\centerline{\includegraphics[width = 0.9\linewidth]{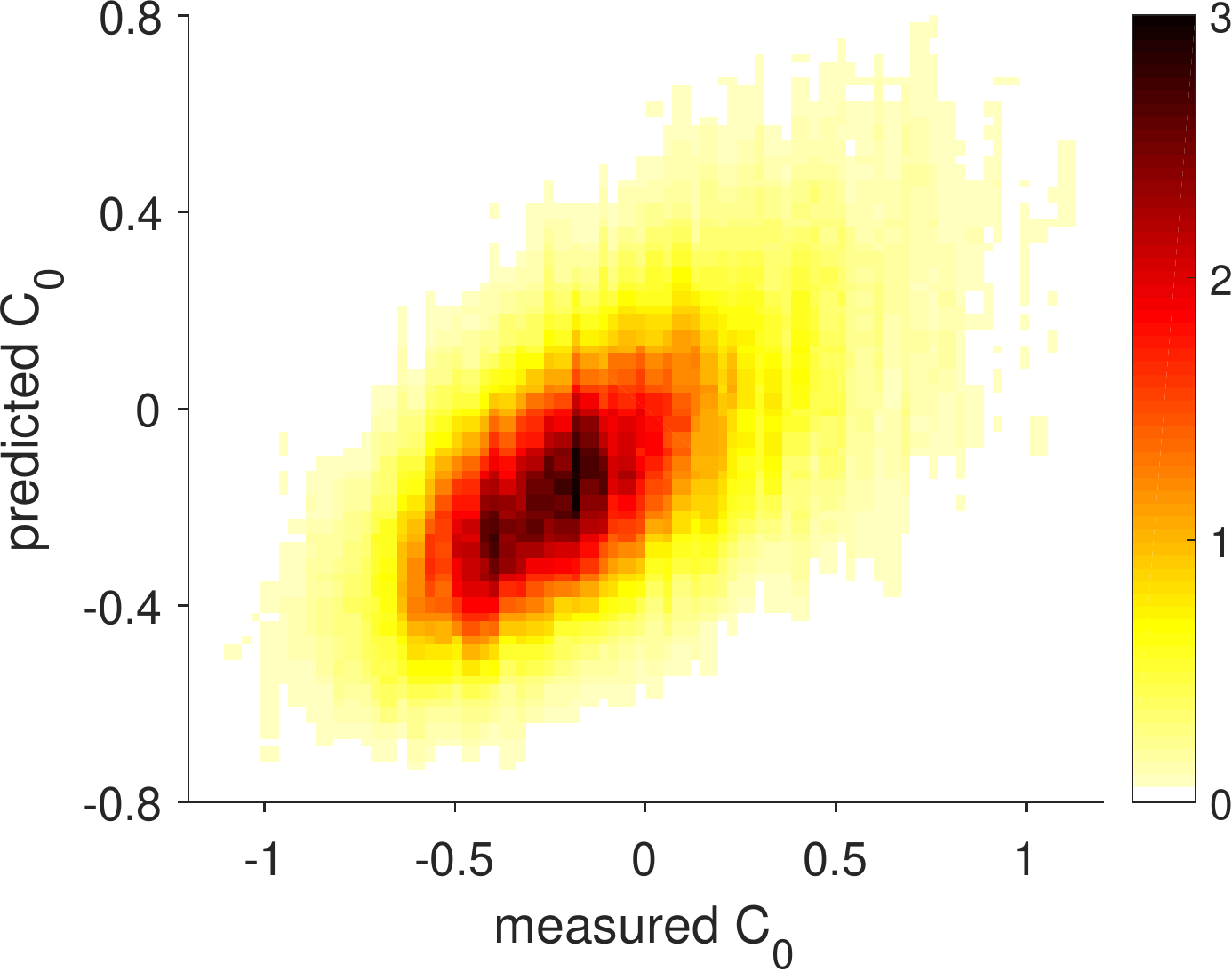}}
\caption{Joint probability distribution of predicted and measured  cyclizability $C_0$ across the ensemble of sequences. \label{prediction}}
\end{figure}

These results suggest that we should take the model in Eq (\ref{2ndModel}) seriously. Again we estimate $ J$ from half of the data, impose translation invariance, and predict $C_0$ for the other half of the data.  Predictions vs measurements are shown in Fig~\ref{prediction} as a joint density; results are obtained from multiple random 50/50 splits into training and testing data. The correlation between predictions and measurements is $r = 0.59\pm0.01$. We can also find the contributions to $r$ from individual modes
\begin{equation}
C_0 = \langle C_0 \rangle + 8\sum_{\rm n} \lambda_{\rm n} f_{\rm n}^2 ;
\label{predictionmodes}
\end{equation}
when all the modes are included, Eq~(\ref{predictionmodes}) reduces  to Eq~(\ref{2ndModel}). Including only the first two modes   results in $r = 0.54\pm0.01$, suggesting that these modes make the largest contribution, as expected from the eigenvalue spectrum, but including all modes provides significantly better predictions.

Should we be satisfied with the quality of predictions in Fig \ref{prediction}, or are we missing something?   We have generated synthetic data on the assumption that the model in Eq (\ref{2ndModel}) is exact, added noise to the resulting values of $C_0$, and repeated our analysis.  In this scenario our ability to recover the underlying model is limited both by the finite number of samples and by the noise level.  With noise levels in the range $\delta C\sim 0.25 - 0.3$ we find the same level of correlation between predictions and measurements as in Fig \ref{prediction}.  There is no direct estimate of the noise level for the measurements in Ref \cite{basu+al_20a}, but with $\delta C\sim 0.25 - 0.3$ we would see a correlation of $r\sim 0.6 - 0.7$ between repeated measurements of $C_0$.  This is slightly smaller than what is found in repeated measurements of the cyclizability on the Cerevisiae Nucleosomal Library \cite{note2}, and comparable to what is seen in comparing random sequences with their reverse \cite{basu+al_20a}.  It thus is possible that the degree of correlation that we see between theory and experiment in Fig \ref{prediction} is close to the limit set by the data itself.

\begin{table}[b]
\begin{tabular}{@{\vrule height 10pt depth4pt  width0pt}c}
\hline\hline            
most cyclizable sequences  \\
\hline     
\tiny{TAAAGGCCCTTTAAGGGCCCTTAAAGGCCCTTTAAGGGCCCTTTAAGGGC} \\
\tiny{AGGGCCCTTAAAGGCCCTTTAAGGGCCCTTAAAGGCCCTTTAAGGGCCCT}\\
\tiny{GCCCTTAAAGGGCCCTTAAAGGCCCTTTAAGGGCCTTTAAAGGCCCTTTA}\\
\tiny{CCTTAAGGGCCCTTAAAGGGCCTTTAAGGGCCCTTTAAGGGCCTTTAAGG} \\
\hline
least cyclizable sequences\\
\hline   
\tiny{CGTCGATCGACGACTGCGACAACGATGATCGTCATCATCATCGATCATCG} \\
\tiny{GATGATCGACGACTGCCGCCATCATCATCGACGTCATCAACGATCGTCGA} \\
\tiny{ATCATCGACGACCGCCGTCATCATCGACGACGACGTTGATCATCGACGAC} \\
\tiny{TCGTCGATCGACGACGGCATCAACGACGATGATCATCATCATCGACGATG} \\
\hline 
\end{tabular}
\caption{Predicted DNA sequences with highest and lowest intrinsic cyclizabilities.} \label{Table1}
\end{table}

What are the sequence features that control DNA flexibility?  Because the eigenvectors are orthonormal, increasing the projection of the sequence onto one eigenvector necessarily decreases the projection onto others.  The largest values of $C_0$ thus are predicted to occur in sequences that have maximal (squared) projection onto the first two modes.
Table~\ref{Table1} shows the four predicted sequences that are extremal in this way. Characteristic features include $5-6$ bp tracts of AT rich segments (i.e. TTAAA, TTTAA, and TTTAAA), followed by $5-6$ bp tracts of CG rich segments (i.e. GGCCC, GGGCC, and GGGCCC), periodically. This  is consistent with previous findings that molecules with AT rich stretches separated by 5 bp from GC rich stretches are more loopable \cite{rosanio+al_15,basu+al_20b}. 
At the opposite extreme, sequences that maximize the squared projection to the last two modes  are predicted to have the smallest values of $C_0$.   These sequences have shorter lengths of repeated nucleotides, and shorter periodicities for the reappearance of the same motifs.

Early work on the sequence dependence of DNA flexibility focused on the influence of dinucleotide pairs, which could be detected in smaller data sets  \cite{sarai+al_89,geggier+vologodskii_10}.  The high throughput experiments of Ref \cite{basu+al_20a} made it possible to see the influence of helical periodicity, leading to models that combine local dinucleotide features across longer distances  \cite{basu+al_20b}.  In many ways our results recapitulate those of Ref \cite{basu+al_20b}, although our model is simpler.

Beyond the analysis of DNA flexibility, our results illustrate the power of correlation functions to extract meaningful information from modern high throughput data.  The analysis is simpler because the experimental sequence ensembles are fully random with no intrinsic correlations, although the discussion can be generalized. It is attractive to see the problem of finding relevant features in DNA sequences as being equivalent to the  problem of finding relevant features in sensory stimuli, where in both cases relevance is defined by some functional behavior of the biological system.

\begin{acknowledgments}
YZ and WB were supported in part by the National Science Foundation through the Center for the Physics of Biological Function (PHY--1734030) and Grant PHY--1607612.  AB was a Simons Foundation Fellow of the Life Sciences Research Foundation, and TH is an Investigator with the Howard Hughes Medical Institute.  \end{acknowledgments}

\appendix

\section{Some details}
\label{details}

Here we give some mathematical details for the analysis of the model in Eq (\ref{2ndModel}),
\begin{equation}
C_0 = \langle C_0\rangle + {1\over 2}\sum_{{\rm kl},\gamma\delta} J_{\rm kl}^{\gamma\delta}( S_{\rm k}^\gamma - 1/4)(S_{\rm l}^\delta - 1/4) .
\label{form2}
\end{equation}
The first thing we notice is that if we shift 
$$J_{\rm kl}^{\gamma\delta} \rightarrow J_{\rm kl}^{\gamma\delta}  +u_{\rm k}^\gamma b_{\rm l} ,$$ 
then we pick up a term in Eq (\ref{form2}) 
$$\sim u_{\rm k}^\gamma b_{\rm l} \sum_\delta (S_{\rm l}^\delta-1/4) = 0.$$ 
This means that, without loss of generality, we can set 
\begin{equation}
\sum_\alpha J_{\rm ij}^{\alpha\beta} = \sum_\beta J_{\rm ij}^{\alpha\beta} = 0.
\label{sumJzero}
\end{equation}
If we think of $J_{\rm ij}^{\alpha\beta}$ as a $(4N)\times (4N)$ matrix, the condition in Eq (\ref{sumJzero}) reduces the rank by $N$, which makes sense since we have $N$ constraints $\sum_\alpha S_{\rm i}^\alpha = 1$.

We look next at the contribution from a   term $J_{\rm kk}^{\gamma\delta}$ that is diagonal in the site indices:
\begin{widetext}
\begin{eqnarray}
\sum_{\gamma\delta}J_{\rm kk}^{\gamma\delta}( S_{\rm k}^\gamma - 1/4)(S_{\rm k}^\delta - 1/4) 
&=& 
\sum_{\gamma\delta}J_{\rm kk}^{\gamma\delta} [\delta_{\gamma\delta} S_{\rm k}^\gamma - (1/4)(S_{\rm k}^\gamma + S_{\rm k}^\delta ) + 1/16]\nonumber\\
&=&
\sum_\gamma 
\left[ J_{\rm kk}^{\gamma\gamma} -(1/2)\sum_\delta J_{\rm kk}^{\gamma\delta} \right] 
S_{\rm k}^\gamma + (1/16)\sum_{\gamma\delta}J_{\rm kk}^{\gamma\delta} \nonumber\\
&=& \sum_\gamma J_{\rm kk}^{\gamma\gamma} S_{\rm k}^\gamma ,
\end{eqnarray}
where in the last step we use Eq (\ref{sumJzero}).  Thus the only site diagonal term that can contribute also is diagonal in the base index, and this contribution  collapses back to a linear model, as in Eq (\ref{PWM}), with $W_{\rm k}^\gamma = J_{\rm kk}^{\gamma\gamma}$.  Thus we can also zero out $J_{\rm kk}^{\gamma\delta}$, since it is redundant. 

Now we are prepared to compute the correlation function that appears in Eq (\ref{C2}),
\begin{eqnarray}
{\bigg\langle} (C_0 - \langle C_0\rangle) ( S_{\rm i}^\alpha -1/4 ) ( S_{\rm j}^\beta -1/4 ) {\bigg\rangle} 
&=& {1\over 2}\sum_{{\rm kl},\gamma\delta} J_{\rm kl}^{\gamma\delta} 
{\bigg\langle}
( S_{\rm i}^\alpha -1/4 ) ( S_{\rm j}^\beta -1/4 )
( S_{\rm k}^\gamma - 1/4)(S_{\rm l}^\delta - 1/4)
{\bigg\rangle} 
\end{eqnarray}
We notice that the average is zero if all the indices $\rm ijkl$ are different; more precisely if ${\rm k}$ is different from all the other indices, then we get zero.  There is no term ${\rm k} = {\rm l}$, so we must have ${\rm k} = {\rm i}$ or ${\rm k } = {\rm j}$; let's try ${\rm k} = {\rm i}$:
\begin{equation}
{\bigg\langle}
( S_{\rm i}^\alpha -1/4 ) ( S_{\rm j}^\beta -1/4 )
( S_{\rm i}^\gamma - 1/4) (S_{\rm l}^\delta - 1/4)
{\bigg\rangle} =
{\bigg\langle}
( S_{\rm j}^\beta -1/4 )(S_{\rm l}^\delta - 1/4)
[ \delta^{\alpha\gamma} S_{\rm i}^\alpha - (1/4)(S_{\rm i}^\alpha  + S_{\rm i}^\gamma ) + 1/16]
{\bigg\rangle} .
\end{equation}
Since ${\rm i} = {\rm k} \neq {\rm l}$, the only remaining choice is whether ${\rm i} = {\rm j}$ or not.  If not, then the average factors,
\begin{eqnarray}
{\bigg\langle}
( S_{\rm j}^\beta -1/4 )(S_{\rm l}^\delta - 1/4)[ \delta^{\alpha\gamma}S_{\rm i}^\alpha - (1/4)(S_{\rm i}^\alpha  + S_{\rm i}^\gamma ) + 1/16]
{\bigg\rangle}
&=& 
{\bigg\langle}
( S_{\rm j}^\beta -1/4 )(S_{\rm l}^\delta - 1/4)
{\bigg\rangle}
[ \delta^{\alpha\gamma}(1/4) - (1/16)]\nonumber\\
&=& \delta_{\rm jl}(1/16) [\delta^{\beta\delta} - (1/4)][ \delta^{\alpha\gamma} - (1/4)] .
\end{eqnarray}
On the other hand, if ${\rm i} = {\rm j} \neq {\rm l}$ we have
\begin{equation}
{\bigg\langle}
( S_{\rm i}^\alpha -1/4 ) ( S_{\rm j}^\beta -1/4 )
( S_{\rm i}^\gamma - 1/4) (S_{\rm l}^\delta - 1/4)
{\bigg\rangle}
=
{\bigg\langle}
( S_{\rm i}^\alpha -1/4 ) ( S_{\rm i}^\beta -1/4 )
( S_{\rm i}^\gamma - 1/4){\bigg\rangle}{\bigg\langle} (S_{\rm l}^\delta - 1/4)
{\bigg\rangle} = 0.
\end{equation}
So what we have shown that there is one term
\begin{equation}
{\bigg\langle}
( S_{\rm i}^\alpha -1/4 ) ( S_{\rm j}^\beta -1/4 )
( S_{\rm k}^\gamma - 1/4)(S_{\rm l}^\delta - 1/4)
{\bigg\rangle}^{(1)} = (1/16)(1-\delta_{\rm ij}) \delta_{\rm jl} \delta_{\rm ki } [\delta^{\beta\delta} - (1/4)][ \delta^{\alpha\gamma} - (1/4)] .
\end{equation}
The other choice was ${\rm k} = {\rm j}$, which we can get by swapping $({\rm i} , \alpha
) \leftrightarrow ({\rm j},\beta)$.  This gives
\begin{equation}
{\bigg\langle}
( S_{\rm i}^\alpha -1/4 ) ( S_{\rm j}^\beta -1/4 )
( S_{\rm k}^\gamma - 1/4)(S_{\rm l}^\delta - 1/4)
{\bigg\rangle}^{(2)} =  (1/16)(1-\delta_{\rm ij}) \delta_{\rm il} \delta_{\rm kj } [\delta^{\alpha\delta} - (1/4)][ \delta^{\beta\gamma} - (1/4)] .
\end{equation}
Putting these together we have
\begin{eqnarray}
\langle C_0 S_{\rm i}^\alpha S_{\rm j}^\beta\rangle_c  &\equiv&{\bigg\langle} (C_0 - \langle C_0\rangle) ( S_{\rm i}^\alpha -1/4 ) ( S_{\rm j}^\beta -1/4 ) {\bigg\rangle} \nonumber\\
&=& {1\over {32}}(1-\delta_{\rm ij}) \sum_{{\rm kl},\gamma\delta} J_{\rm kl}^{\gamma\delta}  \delta_{\rm jl} \delta_{\rm ki } [\delta^{\beta\delta} - (1/4)][ \delta^{\alpha\gamma} - (1/4)] + {1\over {32}}(1-\delta_{\rm ij}) \sum_{{\rm kl},\gamma\delta} J_{\rm kl}^{\gamma\delta}
\delta_{\rm il} \delta_{\rm kj } [\delta^{\alpha\delta} - (1/4)][ \delta^{\beta\gamma} - (1/4)] \nonumber\\
&=& {1\over {32}}(1-\delta_{\rm ij})\sum_{\gamma\delta} J_{\rm ij}^{\gamma\delta} [\delta^{\beta\delta} - (1/4)][ \delta^{\alpha\gamma} - (1/4)] + {1\over {32}}(1-\delta_{\rm ij})\sum_{\gamma\delta} J_{\rm ji}^{\gamma\delta}[\delta^{\alpha\delta} - (1/4)][ \delta^{\beta\gamma} - (1/4)] .
\end{eqnarray}
We recall that $J_{\rm ij}^{\gamma\delta} = J_{\rm ji}^{\delta\gamma}$, so that 
\begin{eqnarray}
\langle C_0 S_{\rm i}^\alpha S_{\rm j}^\beta\rangle_c &=& {1\over {16}}(1-\delta_{\rm ij})\sum_{\gamma\delta} J_{\rm ij}^{\gamma\delta} [\delta^{\beta\delta} - (1/4)][ \delta^{\alpha\gamma} - (1/4)]\nonumber\\
&=& {1\over {16}}(1-\delta_{\rm ij})\left[ J_{\rm ij}^{\alpha\beta} - (1/4)\sum_\delta J_{\rm ij}^{\alpha\delta}
-(1/4)\sum_\gamma J_{\rm ij}^{\gamma\beta} + (1/16)\sum_{\gamma\delta}J_{\rm ij}^{\gamma\delta}\right].
\end{eqnarray}
\end{widetext}
Now we use $\sum_\beta J_{\rm ij}^{\alpha\beta} = 0$, and  our result collapses to
\begin{equation}
\langle C_0 S_{\rm i}^\alpha S_{\rm j}^\beta\rangle_c  = {1\over {16}}(1-\delta_{\rm ij})J_{\rm ij}^{\alpha\beta} = {1\over {16}} J_{\rm ij}^{\alpha\beta} .
\end{equation}
%where in the last step we use $J_{\rm ii}^{\alpha\beta} = 0$.

\begin{figure}
%\begin{subfigure}{.5\textwidth}
%\centering
\includegraphics[width=.9\linewidth]{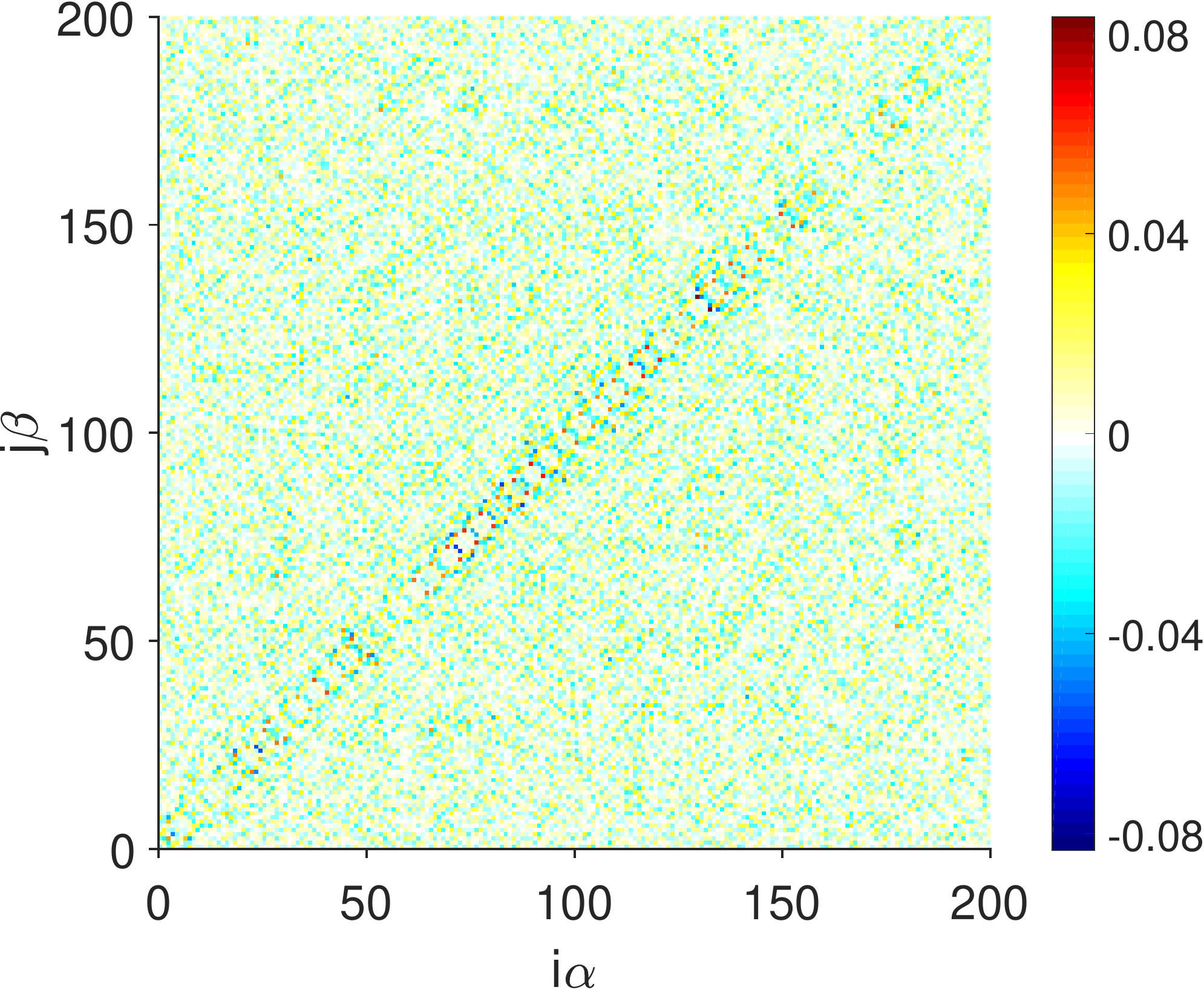}
%\end{subfigure}
\par\bigskip
%\begin{subfigure}{.5\textwidth}
%\centering
\includegraphics[width=.9\linewidth]{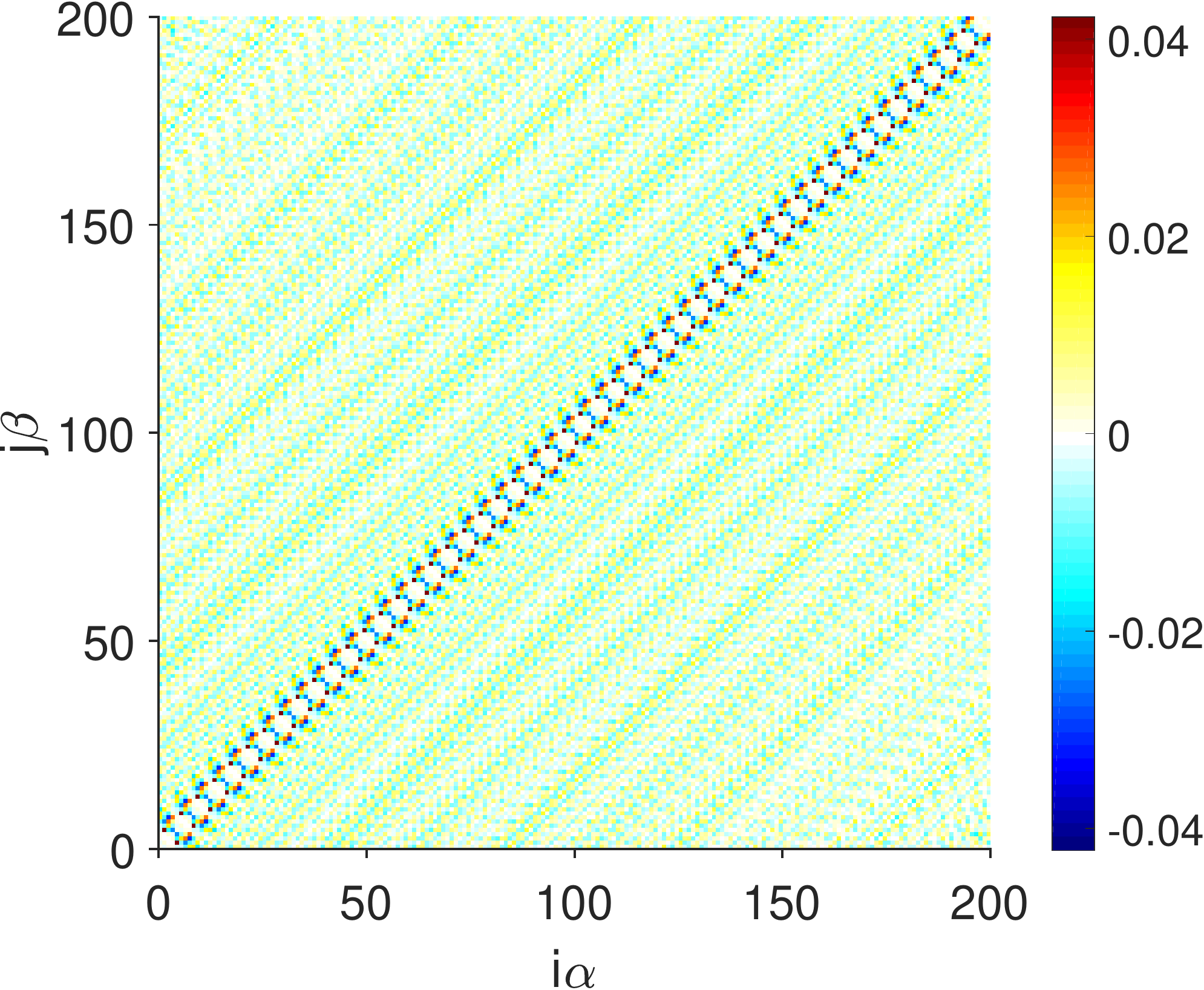}
%\end{subfigure}
\caption{ The interaction matrix $ J$ estimated from the measured correlations [Eq~(\ref{trig_corr})],  before (top) and after (bottom) imposing translation invariance [Eq~(\ref{invariance})]. \label{Jmatrix}}
\end{figure}

\section{Imposing translation invariance}
\label{clean}

We impose translation invariance on the matrix $J_{\rm ij}^{\alpha\beta}$ according to Eq~(\ref{invariance}); Fig~\ref{Jmatrix} shows the $ J$ matrix before and after this treatment.   As noted in the main text,  translation invariance reduces the number of free parameters in $ J$ from $\sim15000$ to $\sim600$
%. By replacing the $J_{\rm ij}^{\alpha\beta}$ with the mean of all possible $J_{\rm i'j'}^{\alpha\beta}$ with $\rm j' - \rm i' =\rm j - \rm i$, we 
and thus raises the signal to noise ratio in the inferred matrix elements. The ``cleaned'' $ J$ not only shows clear stripes near the diagonal, suggesting strong nearest neighbor interactions in determining the cyclizability, but also displays a set of stripes separated at half-helical ($\sim5$ bp) and helical ($\sim10$ bp) period of DNA, suggesting a role more longer ranged interactions in determining DNA flexibility.   

\begin{figure}
%\begin{subfigure}{.5\textwidth}
% \centering
\includegraphics[width=.9\linewidth]{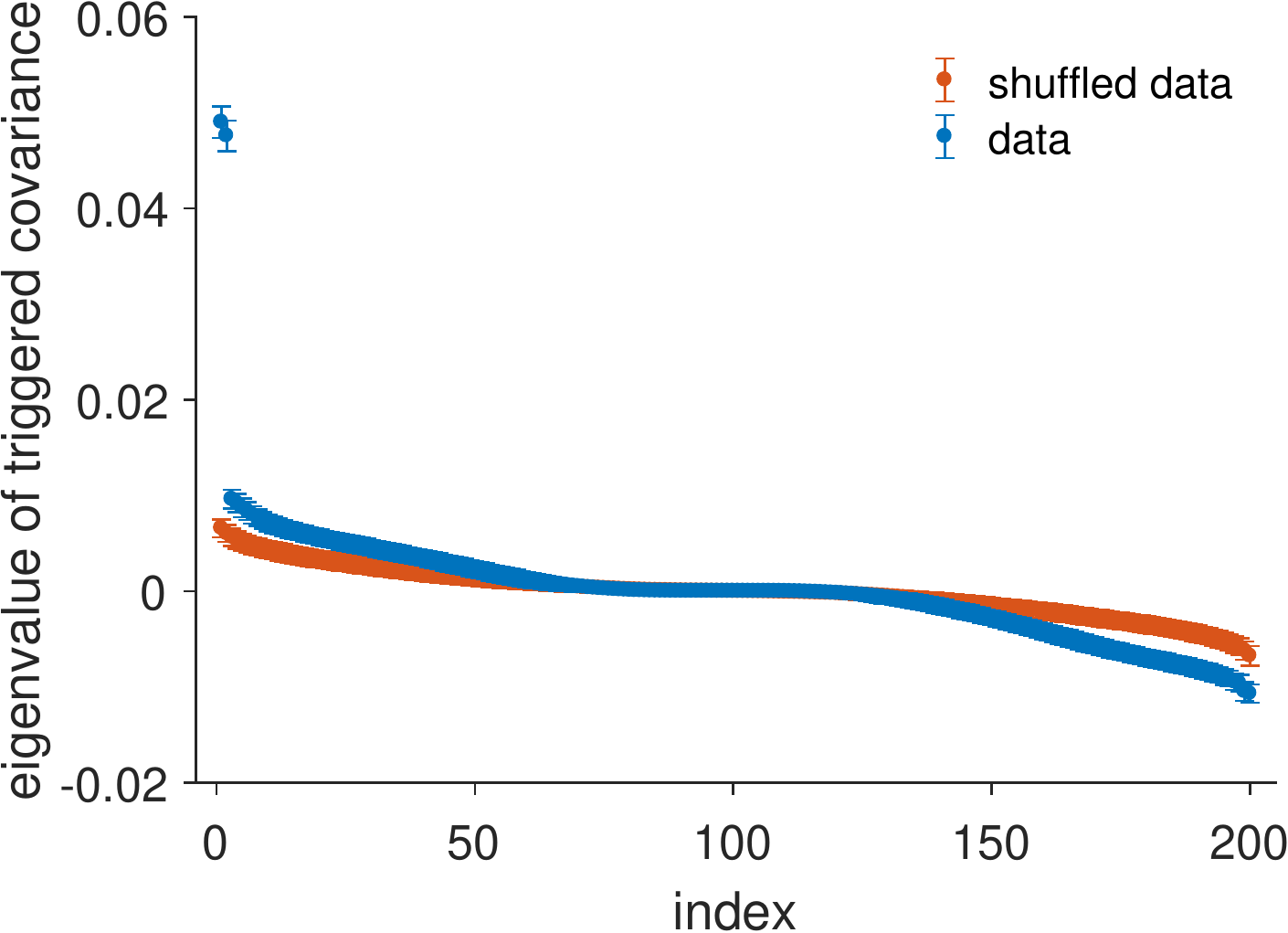}
%\end{subfigure}
\par\bigskip
%\begin{subfigure}{.5\textwidth}
%\centering
\includegraphics[width=.95\linewidth]{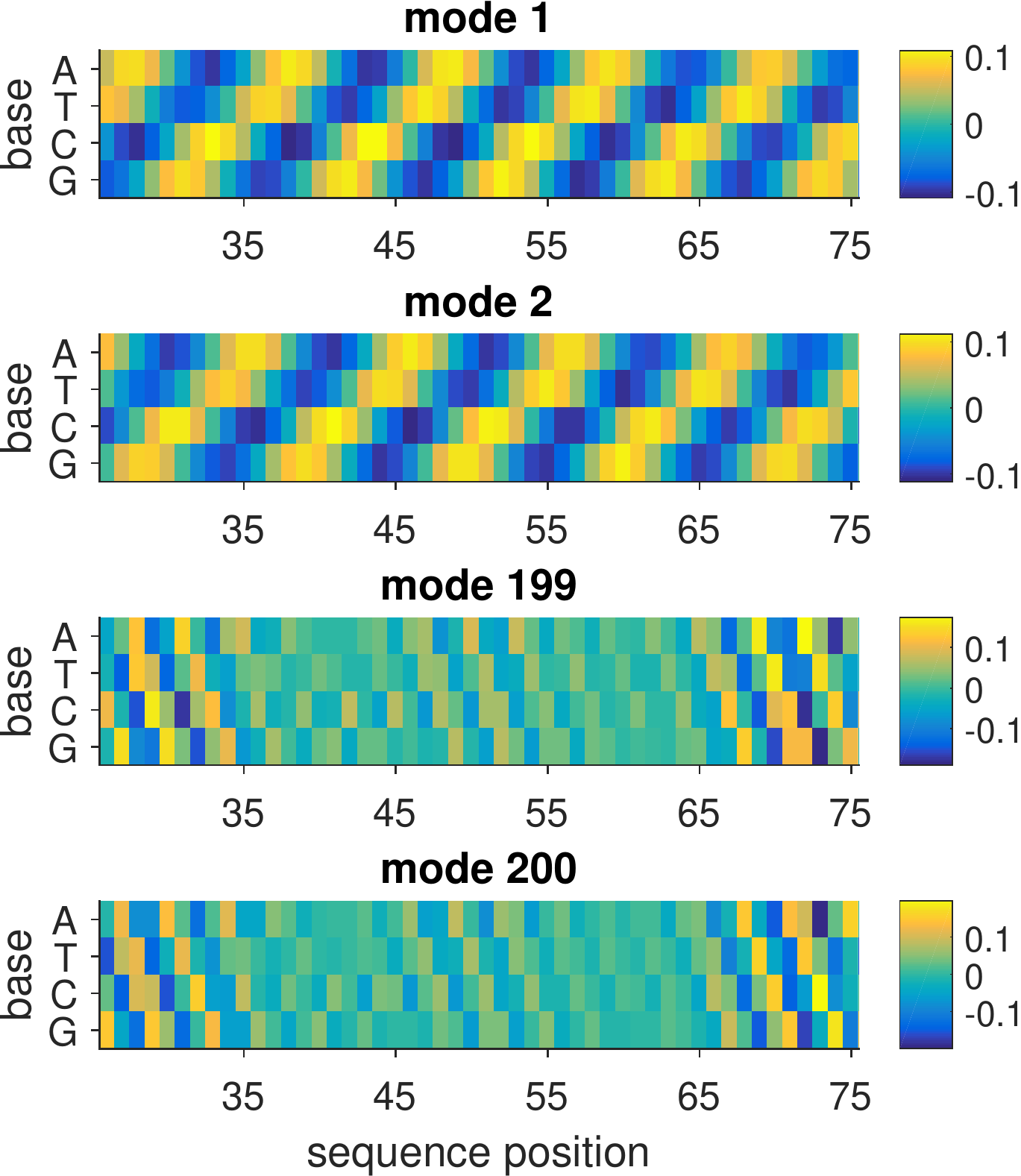}
%\end{subfigure}
\caption{Eigenvalues and leading eigenvectors of the $\langle C_0 S_{\rm i}^\alpha S_{\rm j}^\beta\rangle_c$ after imposing translation invariance. (above) Eigenvalues from real data (blue) compared with results from shuffled data (red); Points are means and error bars are standard deviations across randomly chosen halves of the sequences. (bottom) Eigenvectors $w_{\rm i}^\alpha ({\rm n})$ of the matrix $\langle C_0 S_{\rm i}^\alpha S_{\rm j}^\beta\rangle_c$, for modes with most positive (1, 2) and negative (199, 200) eigenvalues. \label{cleanmodes}}
\end{figure}

The eigenvalues of the cleaned $\langle C_0 S_{\rm i}^\alpha S_{\rm j}^\beta\rangle_c$ ($=J/16$) matrix stand out from the shuffled background with higher signal to noise ratio, both at large positive and large negative values, and the eigenvectors are more clearly periodic.  Results are shown in Fig~\ref{cleanmodes}, which should be compared with Fig \ref{eigenmodes} in the main text.  We note that although the matrix $ J$ is translation invariant, the eigenvectors exhibit clear boundary effects, so that modes 199 and 200 are almost localized at the ends of the sequence.
%We note that the eigenvectors of modes 199 and 200 show some boundary effects, which is due to the translation invariance treatment (similar boundary effects are present in the eigenvectors for shuffled data).

%\textcolor{blue}{From the eigenvectors, we can predict the sequences that maximize and minimize the cyclizability. To do that we require the sequence to maximize $f_{\rm n}^2$ in Eq~(\ref{feat}) for modes ${\rm n}=1,\,2$ (most cyclizable) and for modes ${\rm n}=199,\,200$ (least cyclizable). These sequences are listed in Table~\ref{Table1}.}


\begin{thebibliography}{99}
\bibitem{note1}
We have verified that the maximum and minimum eigenvalues in the shuffled data vary as $1/\sqrt{M}$, where $M$ is the number of sequences in our sample, as expected from random matrix theory.
\end{thebibliography}
\end{document}